# THE BIBLIOMETRIC PROPERTIES OF ARTICLE READERSHIP INFORMATION

Michael J. Kurtz, Guenther Eichhorn, Alberto Accomazzi, Carolyn Grant, Markus Demleitner, Stephen S. Murray, Nathalie Martimbeau and Barbara Elwell

Harvard-Smithsonian Center for Astrophysics, Cambridge, MA 02138

## ABSTRACT

Digital libraries such as the NASA Astrophysics Data System (Kurtz et al. 2004) permit the easy accumulation of a new type of bibliometric measure, the number of electronic accesses ("reads") of individual articles. We explore various aspects of this new measure.

We examine the obsolescence function as measured by actual reads, and show that it can be well fit by the sum of four exponentials with very different time constants. We compare the obsolescence function as measured by readership with the obsolescence function as measured by citations. We find that the citation function is proportional to the sum of two of the components of the readership function. This proves that the normative theory of citation is true in the mean. We further examine in detail the similarities and differences between the citation rate, the readership rate and the total citations for individual articles, and discuss some of the causes.

Using the number of reads as a bibliometric measure for individuals, we introduce the read-cite diagram to provide a two-dimensional view of an individual's scientific productivity. We develop a simple model to account for an individual's reads and cites and use it to show that the position of a person in the read-cite diagram is a function of age, innate productivity, and work history. We show the age biases of both reads and cites, and develop two new bibliometric measures which have substantially less age bias than citations: SumProd, a weighted sum of total citations and the readership rate, intended to show the total productivity of an individual; and Read10, the readership rate for papers published in the last ten years, intended to show an individual's current productivity. We also discuss the effect of normalization (dividing by the number of authors on a paper) on these statistics.

We apply SumProd and Read10 using new, non-parametric techniques to rank and compare different astronomical research organizations

*Subject headings:* digital libraries; bibliometrics; sociology of science; information retrieval

## 1. INTRODUCTION

Digital libraries, such as the NASA Astrophysics Data System (ADS; Kurtz et al. (1993),Kurtz et al. (2000), Kurtz et al. (2004), hereafter Paper1) are able to record detailed information on the readership of individual articles. Essentially these records list who read (accessed) what article when, just as traditional library circulation records list who accessed what book when.

While, as with traditional library records, privacy concerns forbid some potential uses of these data, with proper care they can provide a new and very powerful source for bibliometric measurement. In Paper1 we used the number of times articles were accessed (hereafter "reads") using the ADS as a function of the country originating the query to investigate the worldwide basic research effort.

In this paper we examine the properties of the reads themselves. We compare and contrast their properties with those of citations, and we use these similarities and differences to develop new techniques for the bibliometric evaluation of individuals and organizations.

In section 2 we develop a four component model to describe how the astronomy literature becomes obsolete (as measured by actual article reads) with age. In section 3 we look at the relationship between citations and readership, beginning with the mean relation in section 3.1 and continuing in section 3.2 to look at the relationship for individual articles.

In section 4 we develop a methodology to evaluate the research performance of individuals, using a combination of citations and readership. The read-cite diagram is introduce in section 4.1.1 presenting a two dimensional view of the productivity of individuals, The age-productivity model of section 4.1.2 simply explains the meaning of the diagram.

We develop new productivity measures with differing properties in section 4.2 and we evaluate the age biases of each statistic.

Extending the results for individuals, in section 5 we then develop techniques for comparing organizations.

We conclude in section 6.

## 2. READERSHIP AS A FUNCTION OF AGE

Because the use of the ADS is now the dominant means by which astronomers access the technical literature (see sections 2 and 6 in Paper1) the ADS usage logs can provide a uniquely powerful view of the way an entire discipline (astronomy) uses the technical literature. In this section we will examine the obsolescence (e.g. White and McCain (1989), Line and Sandison (1975), Sandison (1971)) of the technical literature of astronomy as a function of article age based on the actual readership of an article. This is an extension and reexamination of the work explored in Kurtz et al. (2000).

We use, as our basic data source, the log of all article "reads" using the ADS between January 1st and August 20th, 2001. We define a "read" as every time a user who has access to a list of articles, their dates, journal names, titles and authors (such as in figure 1 in Paper1) chooses to view more information about an article. Currently 50% of these "reads" are of the abstract, 38% are of one of the forms of whole text, 8% are of the citation list, and the rest are distributed amongst ten other options. There are more than 4.2 million "reads" in this log.

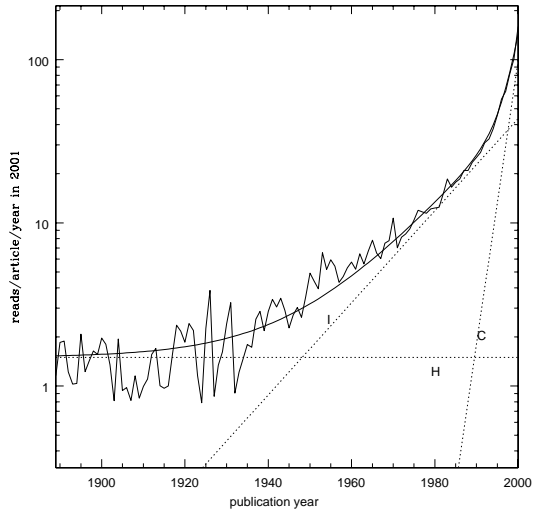

Fig. 1.— The average number of reads per article per year for three U.S. astronomy journals. The thin line is the actual data, the thick line is the model in the text, and the three dotted lines represent the three components of the model.

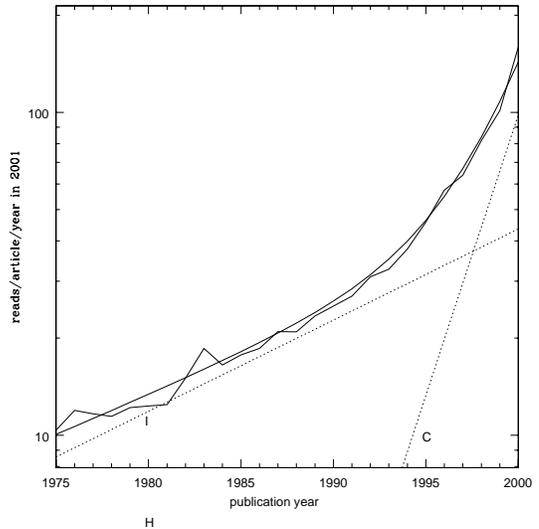

Fig. 2.— An expanded view of Figure 1 showing the most recent 25 years. Note that the model fits the actual data very well.

For this study we extracted reads for the three major U.S. astronomy journals; *The Astrophysical Journal*, *The Astronomical Journal*, and *The Publications of the Astronomical Society of the Pacific*. All three of these journals have been stable over the past century and are currently among the most important astronomy journals. All have had their full text versions beginning with their first issues available on-line through the ADS since well before the beginning of the reporting period. These journals accounted for slightly more than 1.8 million reads in the first 7.66 months of 2001.

### 2.1. *The obsolescence model for reads*

Figure 1 shows the average number of reads per article per year for these three journals as a function of publication year. This shows more than a full century from the first issue of *The Publications of the Astronomical Society of the Pacific* in 1889 through 2000.

Figure 2 shows an expanded view of the last 25 years of data from figure 1. The dotted lines show the relevant three components of the four component readership model of Kurtz et al. (2000), as modified here. In this ad hoc model, research article readership (R) is parameterized by the sum of four exponential functions with very different time constants. We associate these four functions with four different modes of readership: Historical ($R_H$), Interesting ($R_I$), Current ($R_C$) and New ($R_N$). The New ($R_N$) mode, which corresponds to the newly arrived (either on-line or in the mail) issue, cannot be parameterized by the data in figures 1 and 2. The Historical ($R_H$) mode we actually parameterize as a constant, $H_0$, we leave the exponential form in equation 4 (with $k_H = 0$) because other combinations of multiplicative and time constants can also be found which fit the data well, including some combinations with $k_H \neq 0$.

$$R = R_H + R_I + R_C + R_N \qquad (1)$$

where

$$R_H = H_0 e^{-k_H t}$$
$$R_I = I_0 e^{-k_I t}$$
$$R_C = C_0 e^{-k_C t}$$
$$R_N = N_0 e^{-k_N t}$$

and

$$H_0 = 1.5; k_H = 0$$
$$I_0 = 45; k_I = 0.065$$
$$C_0 = 110; k_C = 0.4$$
$$N_0 = 1600; k_N = 16$$

$$t = \text{time since publication in years}$$

The three longer term functions, $R_H$, $R_I$, and $R_C$ are parameterized to fit the data shown in figures 1 and 2. The $R_N$ function is included for completeness; $k_N$ is taken from Kurtz et al. (2000), and $N_0$ obtained by assuming $k_N$ is correct, and ascribing all readership of the *Astrophysical Journal* electronic edition which does not originate with the ADS to the N mode. This is a very crude approximation, but the three component ($R_H$, $R_I$, and $R_C$) model for archival readership is not affected by the N mode usage, which fades very rapidly following publication.

The three mode model is not unique, but does, as figures 1 and 2 show, provide a very good fit to the existing data. No model consisting of only two exponential functions can fit both the recent and historical data, as comparing the two figures makes clear.

Most studies of obsolescence find that the use of the literature declines exponentially with age, and parameterize this with a single number, often called the "half-life," which is related to the coefficient in the exponent by half-life = $log_e(2)/k$, the point where the use of an article drops to half the use of a newly published article. There are several other definitions of half-life in the literature, we use this one. Thus the Historical ($R_H$) component of equation 1 does not have a half-life; the half-life of the Interesting ($R_I$) component is 10.7 years; the half-life of the Current ($R_C$) component is 1.7 years. Kurtz et al. (2000) estimate the half-life of the New ($R_N$) component at 16 days.

Several studies (e.g. Egghe (1993) and references therein) decompose the exponential decay in use into the product of an intrinsic decay and the general growth of the literature. The results presented here are for the mean current use per article published as a function of time since the present, thus we measure directly the intrinsic decay. Kurtz et al. (2000) show the growth of the astronomy literature has been 3.7% per year, measured in terms of number of papers published over the past 22 years.

The total number of papers read over time in each mode is just the integral of the function from zero to infinity, which for a negative exponent is just the ratio of the two constants: $H_0/k_H = \infty$ reads (one and a half reads per year forever); $I_0/k_I = 818$ reads; $C_0/k_C = 275$ reads; $N_0/k_N = 100$ reads. This assumes no growth in the number of reads; if the number of reads per year increases long term at the same rate, 3.7%, at which the number of publications is now increasing, then the constants in the exponents would all be increased by 0.037. This would have very little effect on the integrals of the $R_N$ and $R_C$ functions, but would more than triple the articles read in the $R_I$ mode; and the $R_H$ mode would grow apace with the growth in the number of reads.

### 2.2. *Discussion*

Beginning with Burton and Kebler (1960) there have been a number of studies (see White and McCain (1989) for a review) which suggest that the obsolescence function consists of the sum of two exponentials, which Burton and Kebler (1960) attribute to "classic" and "ephemeral" papers; parameterizations (e.g. Price (1965)) tend to be similar to our $R_H + R_I$ functions.

If we ignore the $R_N$ component, which neither this study, nor any of the other studies of obsolescence could see, we still very clearly find three separate components to the obsolescence function. Why have these three components not been seen until now?

We suggest that the data available to previous studies have not been adequate to see these subtle effects. Most studies have used citation data to determine the obsolescence function. Because it takes time after a paper is published for it to be cited (e.g. section 3) the peak in the $R_C$ mode is obscured in citation data. Also citation studies have substantial problems accounting for the growth of the literature, which has not been at all constant over the past century. Related to this is the determination of the size of the sample universe (the number of relevant papers to the study) at past times.

There are certainly other possibilities, perhaps the obsolescence function is different for reads and cites, and perhaps the very existence of the ADS has changed the way the literature is used. We will explore these questions further in section 3.

The reason why readership studies have not seen the three component nature of archival readership which we see, we suggest, is that the data available in such studies have been too sparse. The largest astronomy library, the Center for Astrophysics Library, has a reshelve rate of about 1,000/month (Coletti 1999) , which is less than 0.2% of the rate of reads in the ADS. Additionally many astronomers keep (and use) their own paper copies of recent journals, which would suppress the $R_C$ mode in library use.

### 3. THE RELATIONSHIP BETWEEN READS AND CITES

Central to bibliometrics is the study of citations (Garfield 1979), and central to the study of citations is the so called normative assumption (Liu 1993) that "the number of times a document is cited ... reflects how much it has been used..." (White and McCain 1989). There have been many articles suggesting problems with citation studies (e.g. MacRoberts and MacRoberts (1989)), and many articles defending them (e.g. Small (1987)). White (2001) and Phelan (1999) discuss these issues.

The readership data discussed in section 2 provide a totally independent, direct new measure of "how much [an article] has been used." Comparing the readership statistics with citation measures will show the similarities and differences between citations, which are an indirect measure of use, but, some would argue, a direct measure of usefulness, and reads, which are a direct measure of use, but perhaps an indirect measure of usefulness. Here we expand considerably on the comparison presented in Kurtz et al. (2000).

### 3.1. *The mean relationship between reads and cites*

While there have been many dozens of studies on obsolescence using citations, and many dozens more using readership as determined by using library circulation statistics (see White and McCain (1989) for review), there are very few studies comparing the two methodologies over the same data. Tsay (1998) compared the readership obsolescence function (obtained by reshelving statistics) for a number of medical journals with the citation obsolescence function for the same journals. He found the half-life of the readership function was significantly shorter than the half-life of the citation function. Tsay (1998) reviewed the literature and found only one previous comparable study; Guitard, in 1985 (discussed in Line (1993)), using photocopy requests as the use proxy, found the citation half-life shorter than the readership half-life.

We have only found two other studies. Cooper and McGregor (1994), also using photocopy data, found citation half-life substantially longer than the use half-life; also they found "no correlation between obsolescence measured by photocopy demand and obsolescence measured by citation frequency." Satariano (1978) used the questionnaire method to find "citation patterns reflect a cross-disciplinary focus that is not found in the journals most often read."

We believe Kurtz et al. (2000) contains the first study using a data-set large enough to show the similarities and differences between the two obsolescence functions. Here we use a substantially improved data-set; we intend this section to supersede the study in section 6.2 of Kurtz et al. (2000).

#### 3.1.1. *Synchronous relation*

Kurtz et al. (2000) found that the instantaneous obsolescence function for articles from the recent technical astronomy literature as measured by citations is simply equal to a proportionality constant times the function measured by reads times an exponential ramp-up to account for the time delay from when an article is first published to when an article which cites that article is published:

$$C = cR(1 - e^{-k_D T}) \qquad (2)$$

where

$$c \approx \frac{1}{20}; k_D = 0.7$$

The proportionality constant, $c$, represents the number of citations per read. This changes with the (always incomplete) citation databases, and with time, as ADS use increases (see section 4 of Paper1). Currently we estimate that the average paper is read about twenty times using the ADS for every time it is cited. In comparing the citation and reads obsolescence functions we have adjusted $c$ to provide the best fit to the samples.

The time delay (parameterized by $k_D$) essentially is caused by inefficiencies in the publication process. Recently Brody (2003) has shown how electronic publication, in particular the use of the ArXiv.org e-print server, has caused this time delay to shorten.

Figure 3 compares the reads and cites obsolescence functions for recent articles. The readership data are for articles from *The Astrophysical Journal, The Astronomical Journal, The Publications of the Astronomical Society of the Pacific*, and *The Monthly Notices of the Royal Astronomical Society* which were read between 1 January 2001 and 20 August 2001. The citation data are taken from those four journals and both *Astronomy & Astrophysics* and *Nature*, where the publication date was also between 1 January 2001 and 20 August 2001. Only citations to one of the four journals in the readership sample were taken; the data contain 45,000 citations.

As can be seen from figure 3 the citation function follows the reads function very closely. In particular the $R_C$ function (section 2.1) clearly has an analog in the citation data, despite the suppression of the steep increase compared with the raw reads due to the exponential ramp-up. The change in slope in the citation function seen beginning about 1994 is exactly what is predicted from the reads function; the number of citations in 1998 and 1999 are more than 40% above that expected by an extrapolation of the exponential decay seen between 1975 and 1990, a decay which corresponds very closely to the $R_I$ function. We suggest this shows that the citation derived obsolescence function has two components with exactly the same parameters as the two mid-range (in time) readership functions.

To examine the obsolescence function over a longer time period we use a different data-set of citations. We take all citations to the four journals in the readership sample from articles published between 1 January 1995 and 20 August 2001 in the ADS database. The data contain 625,000 citations.

We continue to use as our comparison the year 2001 reads sample. Clearly papers published in 1995 could not have cited papers published in 2000, so comparison with recent obsolescence is impossible (this comparison is in figure 3). We use these data exclusively to examine the long term behavior of the obsolescence function.

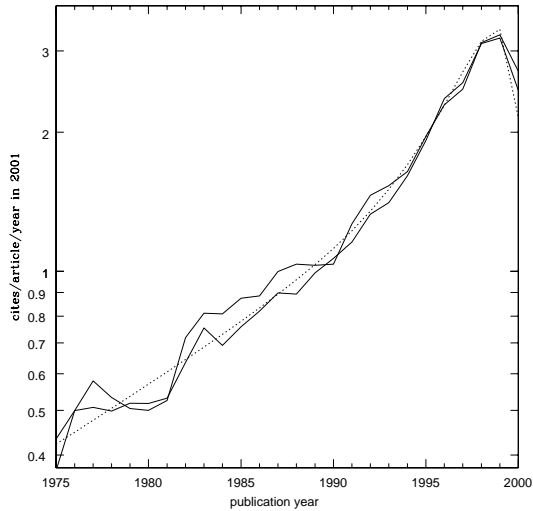

FIG. 3.— A comparison of the $C = cR(1 - e^{-k_D T})$ model with the actual citations from the 2001 sample for papers from the most recent 25 years. The thick line is the actual citations, the thin line is the model, using the actual reads, and the dotted line is the model using the reads model, equation 4.

Figure 4 shows the long term obsolescence function obtained from citation data compared with the readership function. They clearly are **not** the same. The citation function follows the $R_I$ function, but not the $R_H + R_I$ function. This is not a statistical fluke based on having a small number of citations; the number of citations in the period from 1889 to 1940 which are "missing" from the citation function exceeds 5,000. In the year 1900, for example, there were 18 citations in the six year sample, while about 150 would be expected, were the $R_H$ mode to produce citations at the same amplitude as the $R_I$ mode.

We are therefore driven to the conclusion that:

$$C = c(R_C + R_I)(1 - e^{-k_D T}) \qquad (3)$$

where

$$c \approx \frac{1}{20}; k_D = 0.7$$

for research articles in the astronomy literature.

There are a number of possible reasons for the citation obsolescence function to be different from the reads function. There are also a number of possible reasons why the citation obsolescence function measured here does not show the $R_H$ component, whereas this component is seen in other citation based obsolescence functions, beginning with Price (1965). We see no clear candidate explanation which accounts for both differences, however.

### 3.1.2. *Diachronous relation*

As a further test of the citation model in equation 6 we compare the model with the citations to articles published in four different years, 1983, 1986, 1990, and 1992 by the four main journals in astronomy; *The Astrophysical Journal, The Astronomical Journal, The Monthly*

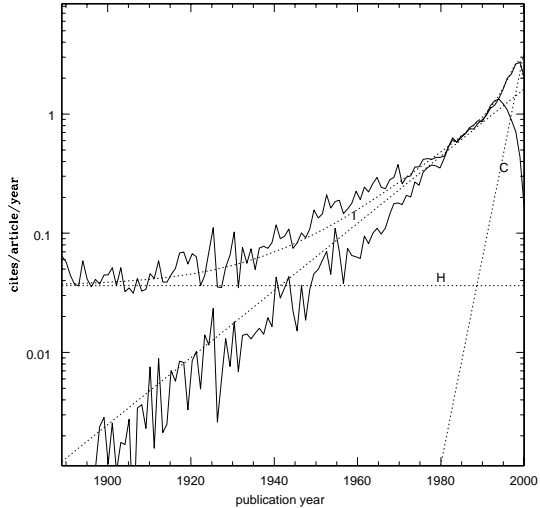

FIG. 4.— A comparison of the $C = cR(1 - e^{-k_D T})$ model with the actual citations for papers from the six year sample for the last 111 years. The thick line is the actual citations, the thin line is the model, using the actual reads, and the dotted lines are the modified reads model from equation 4 and its components.

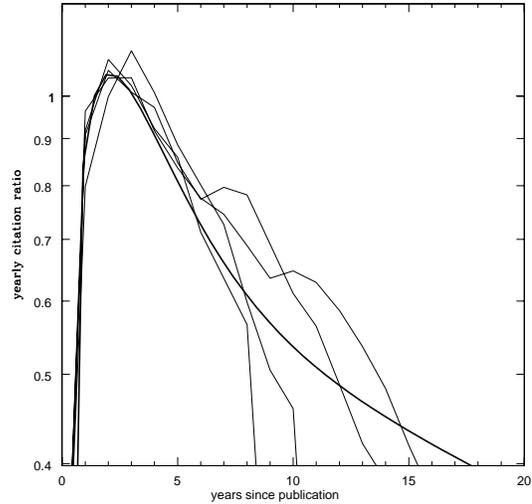

FIG. 5.— A comparison of the $C = c(R_C + R_I)(1 - e^{-k_D T})$ model with the actual citations for papers published in 1983, 1986, 1990 and 1992. The thin lines are the actual data for the four different years, the thick line is the model, using the $R_C$ and $R_I$ components of the reads model of equation 4 and assuming the 3.7% yearly growth rate of Kurtz et al. (2000).

*Notices of the Royal Astronomical Society,* and *Astronomy & Astrophysics.* We use all the citations in the ADS citation database; each year has between 67,000 and 75,000 citations to articles from these four journals.

Figure 5 shows the comparison. The thin lines are the actual number of citations each year to the articles, normalized so that the average citation rate over the first five years after publication is exactly one. The citation rates go to zero when the timespan following publication reaches the present (or more accurately 20 August 2001); thus one can tell which line represents which year, 1992 goes to zero first, 1983 last.

The thick line is the citation model of equation 3, using exactly the same parameters as in equation 1, modified to account for a 3.7% yearly growth in the literature.

It can be noted that the model matches the data as well or better than the curves for the different years match themselves. While these data may be fit with a different citation function, one where the $R_C$ component is zero, the slope of the required $R_I$ function necessary to fit these data is very much too steep to fit the long term decline shown in the data in figure 4. To fit the data in figure 5 with a single $R_I$ function it would have to have a slope $k_I = 0.15$ which is completely excluded by the citation data shown in figure 4. Also the function would then fit only the 1990 data well, the one with the offset peak. We believe the offset peak is caused by a mean three weeks delay in the time of actual publication compared with the publication date in that year, and that the other three years better represent the true function.

Therefore we conclude that this comparison also provides strong evidence for a model where citations follow the function $C = c(R_C + R_I)(1 - e^{-k_D T})$ where the $R_C$ and $R_I$ modes are those useage behaviors seen in the study of the readership data in section 2.1.

### 3.1.3. *Discussion*

We have shown that the citation rate as a function of time is equal to a constant times the sum of two modes of the readership function. There is no *a priori* reason why the constant $c$ in equation 6 should not actually be a function of time. Why should the number of citations per read (about 0.05) be constant, independent of the age of the article?

Examining figures 3 and 4 we see that if $c$ is a function of time it cannot change by more than about 1% per year. This is an extraordinary result, it says that within the (small) measurement error the $C$ function and the $R_C + R_I$ function must be measuring exactly the same thing, the mean usefulness of journal articles as a function of time.

The private act of reading an article entails none of the various sociological influences that the public act of citing an article does (Seglen (1997) lists several of these factors). This suggests that in the mean these factors do not influence the citation rate.

Unless the sum of all the various sociological influences as a function of time is exactly the same as the usefulness of articles as a function of time the existence of these influences would cause $c$ not to be constant. That $c$ is constant means that at every age the total effect of these various influences is zero.

We therefore assert that we have proven that the normative theory of citing (Liu 1993) is true in the mean.

### 3.2. *The relationship for individual papers*

#### 3.2.1. *All articles in the database*

While the citation rate may be a direct measure of usefulness in the mean, there are clearly other factors at play for any individual article. Certainly citations are not the only valid measure of usefulness; newspapers, for example, are very useful but not often cited.

In this section we will compare reads and cites on a per article basis. We believe this is the first time data of this sort have been published. Brody (2003) has recently developed a citation-read correlator for papers in the ArXiv.org e-print database.

Table 1. Reads vs Cites in 2000 for All Articles[a]

| NC\NR | 0 | 1 | 2 | 4 | 8 | 16 | 32 | 64 | 128 | 256 | 512 | 1024 | 2048 | total |
|---|---|---|---|---|---|---|---|---|---|---|---|---|---|---|
| 0 | 20.71 | 17.44 | 16.72 | 16.08 | 15.29 | 14.14 | 12.86 | 11.40 | 9.19 | 5.75 | 0.00 | 0.00 | 0.00 | 2133024 |
| 1 | 12.03 | 11.00 | 11.82 | 12.56 | 12.90 | 12.74 | 11.90 | 10.82 | 9.17 | 6.07 | 0.00 | ⋯ | ⋯ | 36621 |
| 2 | 9.32 | 8.55 | 9.65 | 10.83 | 11.87 | 12.43 | 12.25 | 11.35 | 9.82 | 7.19 | 0.00 | ⋯ | ⋯ | 21418 |
| 4 | 5.81 | 5.21 | 6.30 | 7.83 | 9.40 | 10.95 | 11.68 | 11.39 | 9.95 | 7.64 | 3.46 | 0.00 | ⋯ | 10217 |
| 8 | 2.32 | 2.00 | 3.00 | 3.58 | 5.93 | 7.83 | 9.63 | 10.40 | 9.73 | 7.69 | 4.32 | ⋯ | ⋯ | 3532 |
| 16 | 0.00 | 0.00 | 0.00 | 1.00 | 3.46 | 4.81 | 5.78 | 7.84 | 8.52 | 7.47 | 4.64 | ⋯ | ⋯ | 896 |
| 32 | ⋯ | ⋯ | ⋯ | ⋯ | 1.00 | 2.00 | 3.46 | 4.09 | 5.78 | 6.29 | 4.46 | 0.00 | ⋯ | 190 |
| 64 | ⋯ | ⋯ | ⋯ | ⋯ | ⋯ | ⋯ | 2.00 | 2.00 | 1.58 | 4.09 | 3.58 | 2.00 | ⋯ | 44 |
| 128 | ⋯ | ⋯ | ⋯ | ⋯ | ⋯ | ⋯ | 0.00 | 1.00 | ⋯ | 0.00 | 0.00 | 1.00 | 0.00 | 8 |
| total | 1714338 | 179938 | 112360 | 77346 | 52260 | 32681 | 20256 | 11392 | 4328 | 946 | 94 | 9 | 2 | 2205950 |

[a]In terms of the base 2 logarithm of the cross-tab counts.

Table 1 shows the relationship between reads and cites for all articles in the ADS database. The reads and cites refer to articles read during calendar year 2000 and to articles cited in papers published during calendar year 2000. Table 2 is formatted as follows: the horizontal direction corresponds to the number of reads and the vertical direction shows the number of cites. The data are binned in factors of two, so for example, the sixth column shows the number of citations for articles which were read between 16 and 31 times and the fifth row shows the number of reads for articles which were cited between 8 and 15 times. The actual numbers in the cells are the base 2 logarithm of the actual counts, thus the number of articles which were read between 16 and 31 times and which were cited between 8 and 15 times is $2^{7.83} = 227$.

Looking at table 1 we can see several things, first the vast majority of articles in the ADS database ($10^{20.71} = 77.5\%$) were neither read nor cited during the year 2000. Nearly 97% of all articles in the database were not cited during this period. The most likely number of times which an article was cited was zero for articles which were read fewer than 128 times. 72,926 different articles in the database were cited during this period, while 491,612 different articles were read. The large number of unread articles is because the ADS contains a very large physics data set (it is larger than the astronomy data set); overwhelmingly the ADS is used by astronomers, the physics data are not yet heavily used. At the time these data were captured the ADS did not contain the majority of citations to articles in the physics literature (it does now, however); it did contain most of the citations from the astronomy literature to the physics literature.

Looking at the number of cites (the left column) and finding the most likely number of reads for a given number of cites we find that the most likely number of reads for articles cited zero times is zero, for articles cited 1 time is 8-15, for articles cited 2-3 times is 16-31, etc. The maximum in the readership rate given the citation rate moves very nearly as number of citations times eight. Looking at the cells to the right and left of these maxima we see that the average decline is somewhat less than a factor of two (a difference of one in the base 2 log) for each cell, where each cell is a factor of two in number of reads.

The number of citations are thus a good predictor of the number of reads, with the distribution approximately normal in the log, and with the standard deviation about a factor of two (one in the base 2 log). The same cannot be said of the number of reads, they clearly make a very poor predictor of the number of citations. Looking for example at 64-127 reads the most likely number of citations is zero, but the cell with 2-3 and the cell with 4-7 citations are each close to as full.

With nearly seven times as many different articles being read as cited, the factors which cause someone to read an article are not always the same as the factors which cause someone to cite an article. This is also apparent in that the cell with the most likely number of reads, given a number of cites, contains, in nearly every case, fewer papers than any of the less well cited cells at that readership level.

The most important factor which differentiates reads from cites is youth; as parameterized by the factor $(1 - e^{-k_D T})$ in equation 6. The ratio of reads to cites (in the year 2000, using the ADS) for papers published in 1995 is 17.4, and in 1985 is 17.1. The ratio for papers published in 2000 is 143. 14.5% of all reads in 2000 were to papers published in 2000, but only 2.5% of all cites were to papers published during the calendar year the citing papers were published (2000). 95% of all papers which were read 256 or more times, but which were cited fewer than four times were published within 1.5 years of the end of 2000.

The second most important factor which differentiates reads from cites is the type of article. In 1995 the *Astrophysical Journal*, published 2,258 full length refereed articles. During 2000 2,250 (99.6%) of these were read and 1,541 (68%) were cited. Also in 1995 the *Bulletin of the American Astronomical Society* published 1,272 unrefereed abstracts. During 2000 1,152 (91%) of these were read, but only 22 (1.6%) were cited.

The third factor differentiating cites from reads is the actual content of each article. The most cited and read papers during 2000 make this point. The most cited paper during 2000 (Landolt 1992) is also the second most read paper during this period. Landolt (1992) is a prototypical highly cited paper; it contains the basic calibration data for a fundamental measurement technique.

The most read article during 2000 (Trimble and Aschwanden 2000) was not cited during that time. This is a prototypical newspaper type article, it contains an extensive review of the past year's literature in astrophysics, part of a yearly series of articles by the same authors. While it might be expected that this article would not be cited during the year following its publication it still has never been cited as this present article is written, 2.5 years later, nor have the two articles which followed it or the article which preceded it in the same series (Trimble and Aschwanden 1999, 2001, 2002) each of which was (is) the most read article by ADS users in the year following its publication.

The final factor which differentiates reads from cites is age; the $R_H$ component of readership has no counterpart in the citation function. It is likely that this component of the readership represents papers which are read simply for their historical interest, not because they directly influence a current research problem.

### 3.2.2. The 1990—1997 Astrophysical Journal

Table 2 is similar in format to table 1, but only includes reads and cites during 2000 to papers published in the *Astrophysical Journal* in the years 1990—1997. This eliminates many of the systematic causes of differentiation between reads and cites, such as youth, age, and article type seen in table 1.

The most obvious difference between tables 1 and 2 is that most of the articles represented in table 2 are heavily used, while most of the articles in the whole database are not used at all. 78% of the articles in the whole database were not read in 2000 and 97% were not cited, but only 0.6% of the 1990—1997 *Astrophysical Journal* articles in the database were not read, and 36% were not cited. Recent *Astrophysical Journal* articles are at the core of modern research in astronomy.

The number of cites in table 2 predicts the number of reads to about the same accuracy (a factor of two) as seen in table 1 for the whole database. The number of reads shown in table 2, however, is also a reasonable predictor of the number of cites (also to about a factor of two), which it is not for the whole database.

For the homogeneous sample of standard journal articles used here, this again demonstrates that the normative theory of citing (Liu 1993) is true in the mean; the scatter about this mean is a multiplicative factor of about two, and contains within it all the various sociological, political, and scientific factors which cause variations for individual papers.

Twelve (out of 16,557) of the *Astrophysical Journal* articles in table 3 were cited more times than they were read, 11 were cited once but not read, and 1 was cited twice and read once. This suggests two things. First: ADS users must represent all the subfields of astronomy which are published by the *ApJ*. If there were any large subfield of astronomers who did not use the ADS, the papers they read and reference would show up in table 2 as cited but not read. Second: astronomers overwhelmingly read the articles which they reference. If there were any substantial population of articles which were being *pro forma* cited, but not read they would also appear in table 2.

### 3.2.3. How historical cites influence reads

As shown in section 2 the maximum rate at which articles are read occurs immediately following their publication, and then declines with time. However astronomers learn of the existence of an article during the first year or so following publication, they do not learn of it by a citation, as none exists at that time. It is reasonable to ascribe a causality in this situation: following publication an article is cited because it has been read.

Later in the life of an article this causation is often reversed: an article is read because it has been cited. Pointing readers of an article to other articles of interest to read is one of the principal functions of citations.

Table 3 shows the number of reads during the two year period 2000—2001 for articles published in the *Astrophysical Journal* during the years 1950—1959 versus the **total** number of citations for these articles from the ADS citations database.[1] Very clearly there is a strong correlation of the readership rate of articles with the total number of citations for that article.

The total number of citations can predict the readership rate about as well as the citation rate predicts the readership rate (to within about a factor of two) in the samples of tables 1 and 2. The readership rate is also a predictor of the total number of citations, again to about a factor of two, similar to the ability of the readership rate of 1990—1997 *Astrophysical Journal* articles to predict their current citation rate. Also, as with the 1990—1997 *Astrophysical Journal* articles, the distribution of citations as a function of reads is not a symmetric normal distribution, but is skewed toward fewer citations. Unlike the effect with the whole database (table 1) this does not overwhelm the predictive power of the reads, however.

Although papers with higher total citations tend strongly to be the papers most read, this is not due to the existence of more links to these papers from the on-line reference lists. By examining the reads of *Astrophysical Journal* articles by unique individuals (as determined by their cookie) which follow these individuals accessing the reference list of a different article, as a function of the publication date of the *Astrophysical Journal* article read and the time lag (up to a maximum lag of a couple of minutes) after the reference list was accessed, we can estimate the fraction of articles which were read by directly clicking on a reference link in the back of another article.

For articles less than one year old, this fraction is very near zero, as there has not been time for recently published papers to be referenced; this fraction slowly increases to about 1% for papers ten years old, 2% for papers twenty years old and 3% for papers thirty or more years old. For papers published in the *Astrophysical Journal* between 1950 and 1959 only about 3% of total reads are directly caused by their being cited in on-line reference lists.

Another possibility is that substantial readership of the older literature comes from non-journal based outside links. This is, in fact, true for the second most read

---

TABLE 2. READS VS CITES IN 2000 FOR 1990-1997 *ApJ* Articles[a]

| NC\NR | 0 | 1 | 2 | 4 | 8 | 16 | 32 | 64 | 128 | 256 | 512 | 1024 | 2048 | total |
|---|---|---|---|---|---|---|---|---|---|---|---|---|---|---|
| 0 | 6.64 | 7.17 | 8.89 | 10.17 | 10.79 | 10.61 | 9.28 | 6.66 | 2.81 | ... | ... | ... | ... | 5922 |
| 1 | 3.46 | 4.00 | 6.21 | 8.36 | 9.69 | 10.29 | 9.64 | 7.38 | 3.46 | ... | ... | ... | ... | 3479 |
| 2 | ... | 0.00 | 4.39 | 6.99 | 9.04 | 10.30 | 10.32 | 8.78 | 5.00 | ... | ... | ... | ... | 3688 |
| 4 | ... | ... | ... | 3.91 | 6.52 | 8.84 | 9.96 | 9.38 | 6.70 | 1.58 | ... | 0.00 | ... | 2332 |
| 8 | ... | ... | ... | ... | 1.00 | 5.04 | 7.88 | 8.70 | 7.37 | 3.17 | 0.00 | ... | ... | 863 |
| 16 | ... | ... | ... | ... | ... | 0.00 | 2.58 | 6.04 | 6.79 | 5.32 | ... | ... | ... | 224 |
| 32 | ... | ... | ... | ... | ... | ... | ... | 1.58 | 3.70 | 4.64 | 1.00 | ... | ... | 43 |
| 64 | ... | ... | ... | ... | ... | ... | ... | ... | 0.00 | 1.00 | 1.00 | 0.00 | ... | 6 |
| total | 111 | 161 | 569 | 1622 | 3210 | 4567 | 3927 | 1860 | 444 | 79 | 5 | 2 | 0 | 16557 |

[a] In terms of the base 2 logarithm of the cross-tab counts.

TABLE 3. READS IN 2000+2001 VS TOTAL CITES FOR 1950-1959 *ApJ* Articles[a]

| NC\NR | 0 | 1 | 2 | 4 | 8 | 16 | 32 | 64 | 128 | 256 | 512 | total |
|---|---|---|---|---|---|---|---|---|---|---|---|---|
| 0 | 5.93 | 4.25 | 5.58 | 5.81 | 4.75 | 1.00 | 1.00 | 0.00 | ... | ... | ... | 216 |
| 1 | 3.91 | 2.58 | 3.70 | 5.17 | 5.09 | 2.81 | ... | ... | ... | ... | ... | 111 |
| 2 | 4.32 | 3.70 | 4.86 | 5.83 | 5.83 | 4.32 | 1.58 | ... | ... | ... | ... | 199 |
| 4 | 3.70 | 3.32 | 4.64 | 6.61 | 6.58 | 5.52 | 1.58 | 0.00 | ... | ... | ... | 292 |
| 8 | 2.81 | 3.00 | 3.00 | 6.36 | 6.88 | 6.04 | 3.58 | 0.00 | ... | ... | ... | 302 |
| 16 | ... | ... | 2.32 | 5.09 | 6.34 | 6.44 | 4.39 | 2.32 | ... | ... | ... | 233 |
| 32 | ... | ... | 0.00 | 0.00 | 5.09 | 6.04 | 4.75 | 4.00 | 0.00 | ... | ... | 146 |
| 64 | ... | ... | ... | ... | 2.00 | 4.09 | 4.81 | 4.32 | 2.81 | ... | ... | 76 |
| 128 | ... | ... | ... | ... | ... | 0.00 | 3.00 | 3.17 | 3.17 | 0.00 | ... | 28 |
| 256 | ... | ... | ... | ... | ... | ... | 1.58 | 1.00 | 1.58 | ... | | 8 |
| 512 | ... | ... | ... | ... | ... | ... | ... | ... | ... | 1.00 | | 2 |
| 1024 | ... | ... | ... | ... | ... | ... | ... | ... | ... | 0.00 | | 1 |
| total | 116 | 56 | 129 | 364 | 451 | 312 | 104 | 56 | 19 | 4 | 3 | 1614 |

[a] In terms of the base 2 logarithm of the cross-tab counts.

article in table 2, Allen et al. (1995), which is pointed to by a number of educational web sites, following its appearance in the NASA Astronomy Picture of the Day web site on 1 November 1997 (NASA-APOD 1997).

To check if this practice is widespread among the older material we take the three most cited and most read papers from the 1950's *Astrophysical Journal*, Salpeter (1955), Johnson & Morgan (1953), Abell (1958) and examine the details of their use during the month of September, 2001 using the raw web server logs. These three articles are among the most famous papers in the history of astronomy. If articles are being accessed routinely from outside educational web sites, these articles would be among those so accessed.

We found no indication that outside links play a large role in the use of these articles; two reads of Johnson & Morgan (1953) came from a static link in a Taiwanese web site, and a couple of reads of Abell (1958) came from the NED database (Helou et al. 1995) and another couple from the SIMBAD database (Wenger et al. 2000). The overwhelming majority of reads of these papers came from queries using the ADS system.

In addition to the three very well cited papers we also examined the details of the query types (from the web server logs) for all *Astrophysical Journal* articles published during the 1950's and read during the first ten days of September, 2001. We find the distribution of queries to be consistent with the distribution for the whole database (Eichhorn et al. 2000) with only one clear difference. The number of queries which have a restrictive date range is higher; typical is a query with an author name and an exact year. These exact queries more commonly retrieve highly cited papers than less well cited ones.

### 4. MEASURING INDIVIDUAL RESEARCH PRODUCTIVITY USING READS AND CITES

As was demonstrated in section 3, reads and cites have substantially different properties, but fundamentally measure the same thing, the usefulness of an article. In this section we discuss some of the issues involved in combining measures of cites and reads to assess the scientific productivity of individuals. The possibility that combining information from the access logs of an electronic library with citation information could result in improved productivity measures has been noted before, ( Kaplan and Nelson (2000), Bollen and Luce (2002)).

#### 4.1. *Joint properties of cites and reads for individuals*

To show some of the basic properties of measuring both cites and reads for individuals we use a data-set designed to show the breadth of authors of astronomy research papers; we include old/young, active/inactive, tenured/nontenured, ... individuals in the sample. For each individual we obtain from the ADS databases a list of all his/her papers published before May 2000 and for each of these papers a list of the total number of citations up to May 2000, and a list of the total number of reads in the period between January 1999 and May 2000.

The sample contains 441 individuals whose primary employment is/was in the field of astronomy. 130 of these are members of the astronomy departments of five prominent universities: Princeton, CalTech, University of California at Berkeley, University of California at Santa Cruz, and Cornell. 166 individuals are employed at the Harvard-Smithsonian Center for Astrophysics, including most of the long-term researchers plus everyone listed in the Center's phone book with a Dr. title and whose family name begins with the letters A–G. 53 individuals are winners of the Russell Prize, the highest award of the American Astronomical Society (many of these individuals are deceased, some for more than 40 years). Finally, 92 individuals are all persons who received their PhD in astronomy from a U.S. university during the year 1980, and have published at least one paper in the ADS database since 1990; slightly fewer than half of the persons receiving U.S. PhDs in 1980 are on the list.

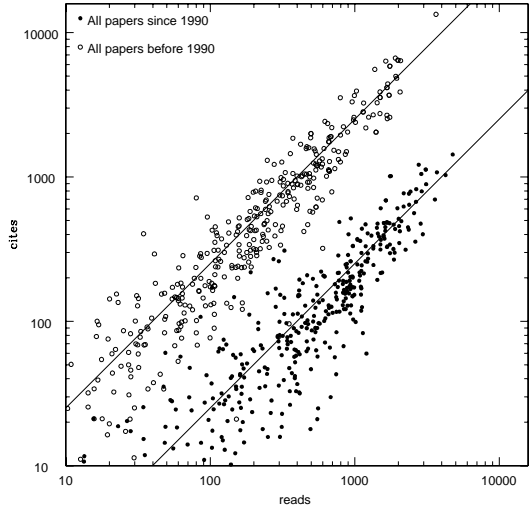

FIG. 6.— Reads vs Cites for individuals in the sample described in the text. Open circles are for papers more than ten years old and filled circles are for papers less than ten years old. The lines show linear relations, with the two relations separated by a factor of ten.

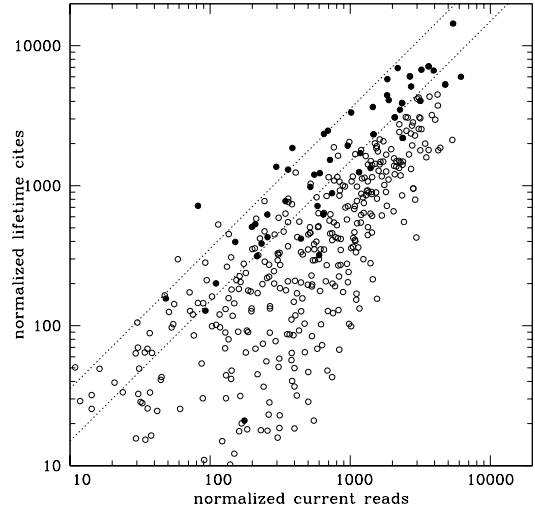

FIG. 7.— Cites vs Reads for individuals in the total sample described in the text. Filled circles are winners of the Russell Prize; open circles are the rest of the sample.

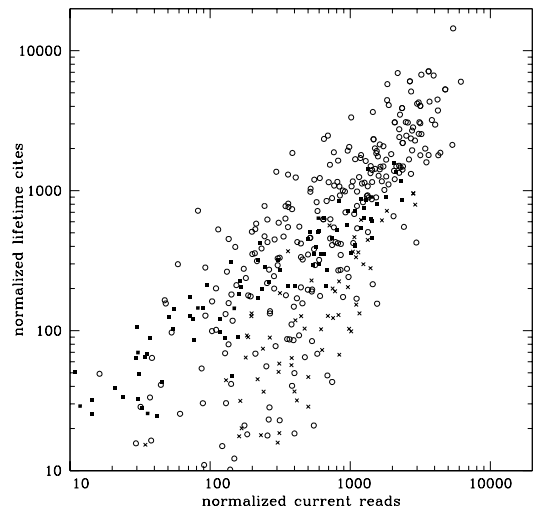

FIG. 8.— Cites vs Reads for individuals in the total sample described in the text. Filled squares are members of the 1980 PhD cohort; crosses are persons who received their PhD after 1985; open circles are the rest of the sample.

#### 4.1.1. The read-cite diagram

Figure 6 shows total cites vs reads for authors from the sample. Each point represents the sum of papers written by a single individual; the open circles for papers published before 1990 and the filled circles for papers written after 1990. Once papers are separated by date the relationship between reads and cites for papers summed over single authors appears quite linear, in agreement with equation 6. The factor of ten difference between the two age groups is fully consistent with that expected using equation 4.

It is obvious from figure 6 that the position of a document in a read-cite diagram is strongly dependent on the age of the document; likewise the position of the sum of an individual author's papers in a read-cite diagram will be strongly influenced by the age of that individual.

Figures 7 and 8 show this effect. Figure 7 shows the total sample in the space of cites vs reads (with the reads and cites normalized for the number of authors on each paper), the filled circles are winners of the Russell Prize; figure 8 is the same, but here the filled squares are the

1980 PhD cohort and the crosses are persons who received their PhD after 1985. The Russell Prize winners are all substantially older than the members of the 1980 cohort; clearly the three distributions form coherent subsamples within the entire sample, and these subsamples are substantially different from each other.

The two dotted lines in figure 7 show the locus for papers which are read once every two (line on bottom-right) or five (line on top-left) years for every time they have been cited. The two dots on the five year line at about 2,500 normalized cites represent W.W. Morgan and M. Schwartschild, two of the most distinguished astronomers of the twentieth century. Each won the Russell prize about forty years ago, and each is currently deceased. Along with table 4 this demonstrates that papers con-tinue to be read as a function of how much they have been cited. The difference in read rates per citation between figure 7 and table 4 is due to the growth of the ADS.

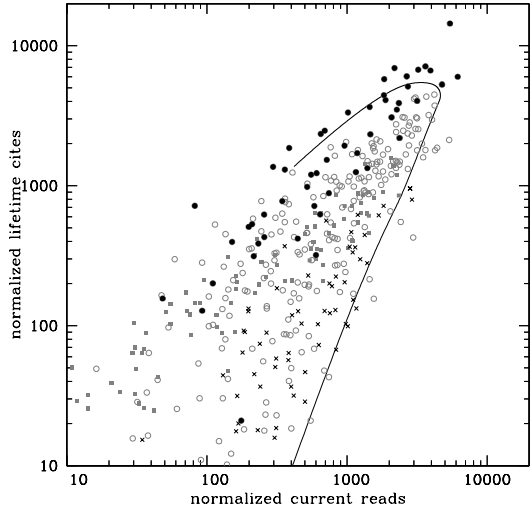

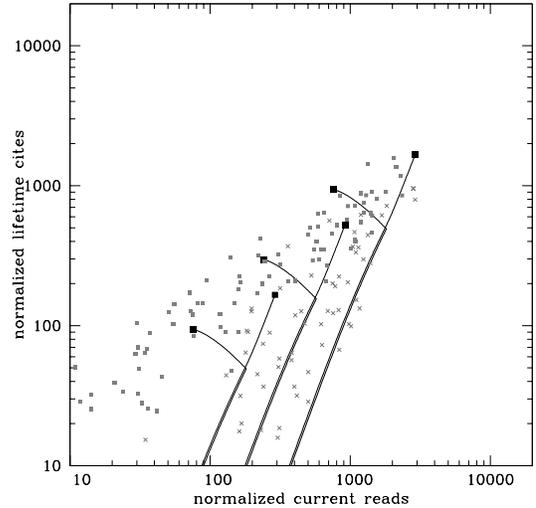

Fig. 9.— Cites vs Reads for individuals in the total sample described in the text. Filled circles are Russell prize winners; the curve represents the model for the current status of persons whose research is at the level of Russell prize winners. The gray squares represent the 1980 PhD cohort, the crosses persons who received their PhD after 1985, and circles the rest of the sample.

Fig. 10.— Cites vs Reads for individuals in a subset of the total sample described in the text. Filled gray squares are members of the 1980 PhD cohort; gray crosses are persons who received their PhD after 1985. The curves represent six different models for current researchers described in the text, the large dark filled squares are the different model predictions for researchers 20 years past the PhD, corresponding with the 1980 cohort.

Another interesting facet of figure 7 is its predictive power. Just to the right of the two year line, at about 3,000 normalized reads and 4,000 normalized cites is a group of three points, one a filled circle representing a Russell prize winner. One of the two open circles in the triplet represents W.L.W. Sargent; Professor Sargent won the Russell prize six months after this plot was made.

#### 4.1.2.  The age-productivity model

Using the relation between reads and cites of equation 6 and the fact that older papers are read according to how much they have been cited, we can understand these diagrams in detail. The right and top outer envelopes of figures 8 and 7 represent the most productive astronomers at different ages. The basic shape of an individual's career tract in this diagram is similar to a (somewhat tilted) figure 7, with the sharp angle appearing at the time of retirement.

A simple, plausible, but non-unique model which fits the data is that a person starts off following the PhD being productive, this productivity grows by about a factor of two during the first seven years of one's career and stays stable for the next twenty-three years. At thirty years past the PhD one's productivity begins to decline until forty-two years past the PhD when one is fully retired and the number of reads declines to one per citation per couple of years.

Figure 9 shows this model for the current reads vs lifetime cites of individuals currently between one and one hundred years past the PhD. The curve bends down for individuals more than 45 years past the PhD because when they were active astronomy was a smaller field than currently, as discussed in section 3.1.2.

The fact that the large majority of Russell prize winners lie near the curve suggests that within small factors the model is an accurate representation of the productivity paths of the most productive astronomers. The tightness of the relation for Russell prize winners is underestimated in the actual data (the filled circles) due to the increasing incompleteness of the citation data base as a function of age.

Figure 10 shows how the age-productivity model can acccount for the distribution of the 1980 PhD cohort in the read-cite diagram. The gray symbols are the same as in figure 9, the small gray squares are the 1980 cohort and the gray crosses are persons now working as astronomers who received their PhD after 1985. The solid lines are based on the model in figure 9. There are three groups of two models, these groups are separated from each other by equal multiplicative factors of $\sqrt{10}$ in total productivity.

The pairs of curves at each of three productivities represent the first twenty years of two different extreme career paths for individuals. The taller of the two curves represents the same model career as the curve in figure 9, the first twenty years of the career of a person who works at his/her full capacity; the shorter curve (the one with the large "hook") represents the career of someone with exactly the same latent productivity as the adjacent taller curve, but who stops performing astronomy research after ten years (perhaps leaving the field).

The six large filled black squares represent a person 20 years past the PhD (corresponding with the 1980 cohort) for each of the six models (three latent productivities, with the most productive being a factor of ten times more productive than the least productive, times two different career paths). The large black squares have essentially the same distribution as the small gray squares, thus this simple model is able to account for the distribution of the 1980 cohort in a direct way.

There are some small gray squares, those on the lower left in the diagram, which are not accounted for by the six model points. These are the least cited/read individuals. They can be accounted for either by ascribing to them a lower latent productivity, or an earlier time when they left active research. The spread in latent productivity of active researchers who got their PhDs after 1985 (the gray crosses) is about a factor of ten, and is well contained within the model lines in figure 10, thus we postulate that the natural variation of latent productivity in astronomers is about a factor of ten, and is not great enough to account for the very low cited/read individuals, thus they must have essentially left active research before ten years past the PhD. Although all in the 1980 cohort were chosen to have been co-author on at least one article of any kind since 1990 (which would be ten years past the PhD) examination of the publication records of the least cited/read confirms that they effectively stopped doing research before then.

### 4.1.3. *Discussion*

The read-cite diagram is a powerful new method for examining the research productivity of individuals; by comparing both measures one is able to see differences which would not be visible using either reads or cites alone. For example, if we take two hypothetical researchers, A and B, both A and B are the same age past the PhD, and both A and B have the exact same total productivity from their PhDs to now. Person A began more productive than B, and has become less productive than B over time; person B began less productive than A and has become more productive. Because citation measures more heavily weight past work, and readership measures more heavily weight recent work person A would have more citations than person B, while person B would have more reads than person A.

Combined with the use of similar cohorts, especially same age cohorts, for comparisons, the read-cite diagram allows an individual's research to be evaluated in a manner which is more accurate and fair than is possible with the use of citation counts alone. It does not, however, provide a measure accurate enough to preclude the necessity of careful and direct examination of the record of the person being evaluated.

### 4.2. *New productivity measures for individuals*

It is often useful to have a single number to describe the productivity of an individual, particularly when aggregations of individuals are being studied. Citations (Garfield 1979) have served this role almost exclusively. The existence of the readership information allows other measures to be developed which in many cases are better suited to the particular study being done.

To investigate the properties of different productivity measures we have created a data-set containing the readership information for the calendar years 2000 and 2001, and the lifetime citation information for articles published before 1 January 2002 for a set of astronomers which is all tenured faculty in the ten highest rated astronomy departments in the United States (National Research Council 1995) and the federal civil servant astronomers at the Smithsonian Astrophysical Observatory, arguably the leading U.S. federal laboratory for as-tronomical research. For each of these 308 individuals we have also found the year their PhD was granted. We will call this data set the tenured faculty sample.

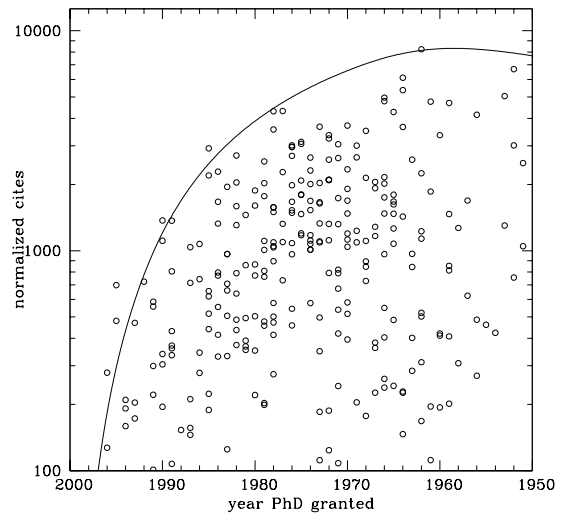

Fig. 11.— Total normalized cites for tenured faculty from the leading astronomical organizations vs. the year each person received his/her PhD. The solid line is the cites model described in the text.

Figure 11 shows the normalized (by dividing by the number of authors for each paper) citations vs PhD date for individuals in the tenured faculty sample. The solid line represents the citation component of the individual productivity model discussed in section 4.1.2, it seems to match the outer envelope of the data quite well.

Figure 11 shows clearly the most important drawback of using citations as a productivity measure: they are very strongly biased towards older individuals. Of the top ten astronomers in figure 11, one received his PhD in the 1940s, three in the 1950s, and six in the 1960s. Using the solid line as a line of constant ability shows that a person twenty years past the PhD (*i.e.* at peak productivity) will only have half the citations of a similar individual forty years past the PhD, (*i.e.* at retirement).

Another measure of productivity is the rate at which an individual's papers are being read. Figure 12 shows the normalized read rate (for the two year period) for members of the tenured faculty sample vs. PhD date. The solid line represents the readership component of the model in section 4.1.2.

Compared to the citation information the reads are much less biased as a function of age. They are, however, still biased, but toward younger individuals. Of the top ten astronomers in figure 12 one received his PhD in the 1960s, three in the 1970s, four in the 1980s, and two in the 1990s. Only one person (J.P. Ostriker) is in both lists.

Just as citations have been used with great profit in a large number of studies, the readership information, which is a more direct measure of current usefulness, seems also well suited for use in bibliometric studies.

### 4.2.1. *The SumProd statistic*

While the direct use of the readership information may eliminate much of the age bias inherent in citation anal-ysis it introduces other difficulties. As can be seen in figure 2 and in equation 4 the readership of an article declines very rapidly following its publication. Thus, for example, since individuals tend to have not more than a few very highly read/cited papers in their careers, the read statistic will be substantially elevated if the sample period coincides with the release of such a paper compared to a few years before or later. Citations are much more stable to these influences.

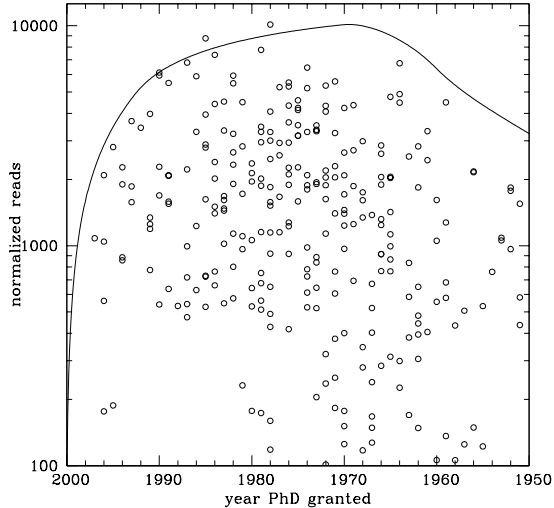

FIG. 12.— Total normalized reads in 2000 and 2001 for tenured faculty from the leading astronomical institutions vs. the year each person received his/her PhD. The solid line is the reads model described in the text.

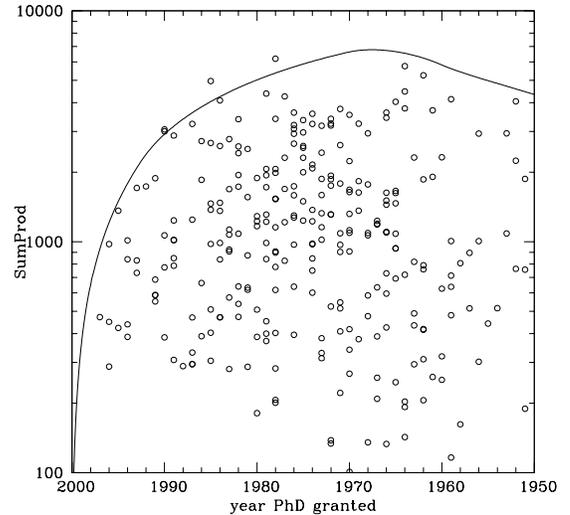

FIG. 13.— The SumProd statistic for tenured faculty from the leading astronomical institutions vs. the year each person received his/her PhD. The solid line is the productivity model described in the text.

It seems justified to create a statistic which has the attributes of both the reads and the cites. We make one by adding the scaled reads to the cites, and name the statistic SumProd. $SumProd \equiv (f \cdot reads + cites)/2$, where the reads and cites are normalized by the number of authors in each paper, and the reads are multiplied by the factor $f$ to make the numerical weight of the current readership rate equal to the lifetime sum of the cites, over the entire sample. Dividing by two makes the numerical range the same as for citations. Figure 13 shows this statistic for the tenured faculty sample; the solid line is derived from the sum of the reads and cites lines in figures 11 and 12. The scaling factor for the reads was 0.8.

Because SumProd uses two measures with substantially different age dependencies, it is substantially more stable for younger scientists than citation counts, and it is also more stable for older scientists than the readership rate.

Of the top ten astronomers in figure 13 two received their PhDs in the 1950s, three in the 1960s, three in the 1970s, and two in the 1980s. SumProd has substantially less age bias than citation counts alone; for nearly all studies where citation counts are an appropriate measure, we suggest that SumProd would be a more appropriate measure.

For studies where age bias must be minimized, one can perform a model dependent age normalization of the data, by dividing by a model of productivity vs age, such as shown in figures 11, 12, and 13. Figure 14 shows the SumProd statistic from figure 13 divided by the age-productivity model in that figure, that is, the age normalized relative productivities of the members of the tenured faculty sample.

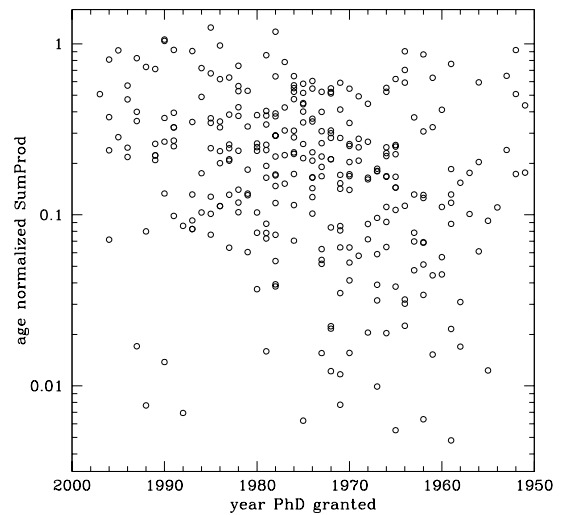

FIG. 14.— The age normalized SumProd statistic for tenured faculty from the leading astronomical institutions vs. the year each person received his/her PhD.

### 4.2.2. The Read10 statistic

SumProd is intended to measure an individual's lifetime scientific production, where for individual's who are still active, this involves an extrapolation of their current status. Once an individual has written a paper, SumProd counts the total citations and current reads; it is not a measure of current activity. Indeed an individuals SumProd measure remains nearly constant in perpetuity. For example, W.W. Morgan would rank in the top quarter of the tenured faculty sample according to SumProd. Professor Morgan wrote his last scientific paper about twenty years ago, and died in 1994.

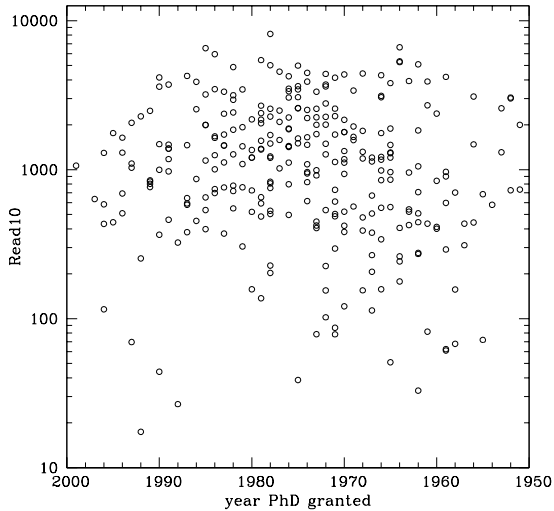

Fig. 15.— The Read10 statistic for tenured faculty from the leading astronomical institutions vs. the year each person received his/her PhD.

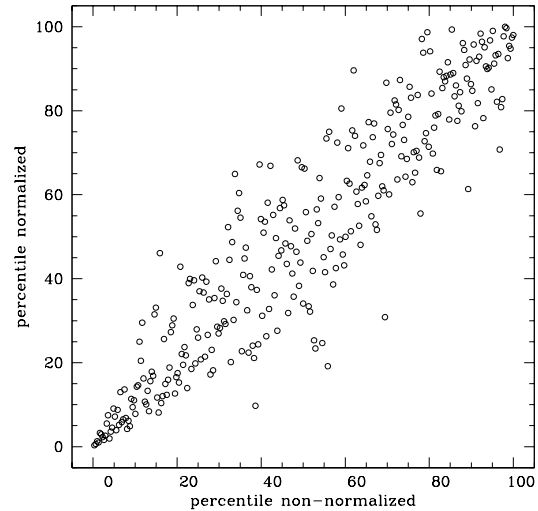

Fig. 16.— The percentile rank of members of the tenured faculty sample obtained by using a version of the SumProd statistic which has not been normalized for number of authors vs. their percentile rank obtained using the ordinary (normalized) SumProd.

We create the Read10 statistic to measure the current activity of individuals. Read10 is the current readership rate for all an individual's papers published in the most recent ten years, normalized for number of authors. Figure 15 shows the distribution of Read10 as a function of an individual's age. The only obvious age bias is for researchers who have not yet been in the field ten years.

For persons in the first ten years of their careers Read10 is exactly the same as the normal readership measure of figure 12. Read10 is very sensitive to changes in output; if one stops publishing Read10 will drop to half its value in about three years, and drop to zero in ten years.

Read10 vs age diagrams, such as figure 15, can provide a powerful means of analyzing the staffing practices of an organization over time. If in some places professors essentially "retire" from research following tenure, this effect would be immediately obvious in these diagrams; likewise if more recent hires are not performing at the level of their elder collegues it would also be obvious.

### 4.3. *The effect of normalization*

All the statistical measures developed in this section have used readership and citation measures which have been normalized to account for the number of authors of each paper. This assumes that each author bears equal responsibility in the creation of a paper.

This assumption is least accurate when applied to a single paper; it becomes more accurate when applied to all the papers of an individual author, as is done in all the plots in this section, and becomes a quite reasonable assumption when applied to groups of individuals.

The alternative measure, ascribing the total reads and/or citations for each paper to each author is clearly not viable for studies of groups of individuals, as the importance of a citation or read would depend on the number of authors of a particular paper who are members of a particular group.

Certainly many analyses of citation counts for individuals are performed with the raw (not normalized by number of authors) counts. These studies can often come to different conclusions than studies using normalized counts would; Herbstein (1993) has argued that the normalized counts are more accurate. These differences are becoming more pronounced as the number of authors per papers increases, (e.g. Schulman et al. (1997)), and especially due to the trend to have papers with very long author lists.

Figure 16 shows this effect for members of the tenured faculty sample. We created a non-normalized analog of SumProd and sorted the list of people by their non-normalized score. This we compared with their rank using the ordinary, normalized SumProd. The ranks are scaled into percentiles, 0–100. While there is a basic trend, that a person who is low/high in one statistic will be low/high in the other, there is a large scatter. It is equally likely that a person at the median of one statistic will be anywhere from the 30th to the 70th percentile of the other. There is an individual who is in the 97th percentile using the non-normalized statistic who is only at the 70th percentile using the normalized statistic. None of the top five individuals in one statistic overlap with the top five in the other.

## 5. MEASURING THE PRESTIGE AND PRODUCTIVITY OF RESEARCH ORGANIZATIONS

Bibliographic measures of scientific productivity have long been tools to assess various aspects of an organization's stature, progress, and desirability. Ranking similar research organizations relative to each other, either by reputations (e.g. National Research Council (1995), hereafter NRC95) or using bibliographic measures, is often used as a tool in determining public policy.

Additionally bibliometric measures are often used by internal review committees to assess the personnel policy of an organization or department.

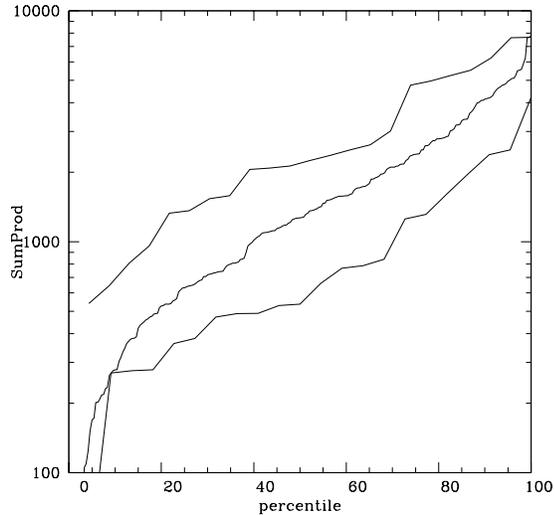

FIG. 17.— The percentile rank of members of the ten faculty sample using the SumProd statistic (thick line) and for members of the Princeton (top thin line) and Cornell (bottom thin line) faculties.

In this section we develop techniques to analyze the performance of groups of individuals, using the bibliometric measures developed in the previous section.

### 5.1. Productivity–Percentile Diagrams

Given a productivity measure or score for individuals, whether SumProd, Read10, number of citations, or some other, how can one compare groups of individuals, such as university departments? Traditional methods are to use the sum or average of the individual scores. Because of the logarithmic nature of these measures this is not a satisfactory technique; for example Alan Sandage, the most cited living astronomer, has a normalized citation score about twenty times the median score for the tenured faculty sample. As a large astronomy department faculty has about twenty members, it is obvious that one or two very high scoring individuals would skew any sum or average.

We develop a technique which compares the distribution of rank ordered individual scores for an entire faculty with the distribution of rank ordered scores for members of several faculties together; figure 17 illustrates the technique.

To compare the top astronomy departments in the U.S., we create a sample which is comprised of the 240 tenured faculty of the top ten astronomy departments according to NRC95, this sample (which we name the ten faculty sample) is the same as the tenured faculty sample of section 4.2, but with the Smithsonian federal employees removed. We then rank order all these individuals, and scale their ranks onto a 0–100 percentile scale. The thick line in figure 17 shows this distribution, using SumProd as the measure.

To compare an organization with the ten faculty sample, we make the same score-percentile distribution for the persons within that organization. The thin lines above and below the thick line in figure 17 show this distribution for the Princeton and Cornell faculties. Next we (numerically) shift the score-percentile distribution for the reference sample up or down to achieve the best fit to the distribution for the organization being compared. As this is a log plot, this is equivalent to multiplying the productivity score of each person in an organization by a constant factor to equal the productivity of the reference sample at that person's percentile rank, and taking the average factor for all individuals in the organization as the result. We only use individuals above the 20th percentile in their organizations in order to minimize the effect of the lowest performers.

The multiplicative factor for Princeton in figure 17 is 1.70 and for Cornell is 0.52. Table 4 shows the scores and rankings for the top 11 organizations as ranked by NRC95, based on perceptions of faculty quality. Both SumProd and Read10 scores and rankings are shown, along with the NRC95 ranking. By construction the ten faculty sample has a multiplicative factor of 1.00 (to match itself).

The members of each faculty were simply taken from each department's web page, taking the professors and associate professors, with two exceptions. M.I.T. does not have an astronomy department, we used the faculty (from four departments) associated with the Center for Space Research. The NRC95 study ranked Harvard, however Harvard is a part of the Harvard-Smithsonian Center for Astrophysics, and it is essentially impossible to separate the organizations from the outside. We have taken as a sample closest to an academic department the persons listed as professors or lecturers on the Harvard web page, 75% of these individuals are Smithsonian federal scientists.

All three rankings are similar; in particular the top four institutions are the same in each ranking, with Princeton, CalTech, and Harvard-Smithsonian each getting one 1st ranking, and U.C. Berkeley being third in all three measures. The only important difference between the SumProd and Read10 rankings and the NRC95 ranking is that the U. Hawaii is higher in both SumProd and Read10, and the U. Chicago is lower.

The NRC95 study ranked more than two dozen astronomy faculties, some of these would doubtless have Read10 or SumProd scores higher than some of the eleven in table 4. The University of Minnesota, for example was ranked 24th by NRC95, but would be ranked eighth by either SumProd or Read10. U. Minnesota has a distribution which is substantially flatter than the reference distribution, and is between one half and one third the size of the eleven organizations ranked. A complete analysis of the astronomy departments in the United States is beyond the scope of this article.

The productivity percentile method measures a type of average productivity for an entire faculty; it is sensitive to the exact composition of the membership of the organizations being ranked. In table 4 we have attempted to keep the samples uniform, by restricting the membership to tenured professors. Much astronomy research in the United States is performed in government supported laboratories, and the employees of these organizations cannot directly be compared with the all faculty sample.

Perhaps the two largest federally supported organizations which perform astronomical research in the U.S. are the Smithsonian Astrophysical Observatory and the Goddard Space Flight Center, which has two laboratories for astronomy: The Laboratory for High Energy Astro-physics, and The Laboratory for Astronomy and Solar Physics. The number of federal civil servant scientists (the closest approximation to academic researchers) at SAO is 64, about two or three times as large as the astronomy departments in table 4, the number of civil servant scientists in the combined labs at Goddard is 100.

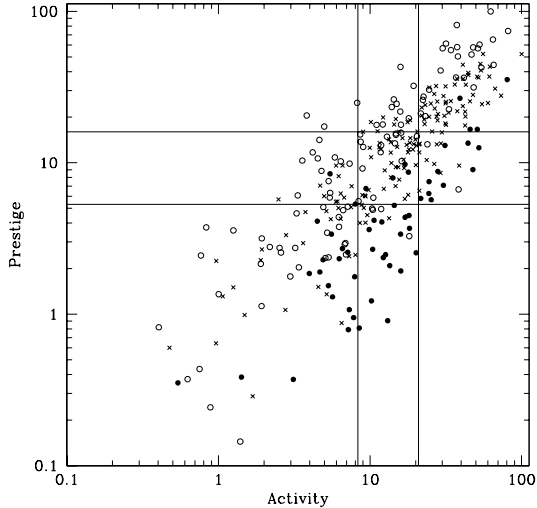

Fig. 18.— The score in percent of the maximum score of members of the tenured faculty sample using the normalized citation statistic (Prestige) vs. their score in percent of maximum using the Read10 (Activity) statistic. Circles are for PhD year before 1970, crosses between 1970 and 1985, and dots after 1985. The lines divide the sample into equal thirds.

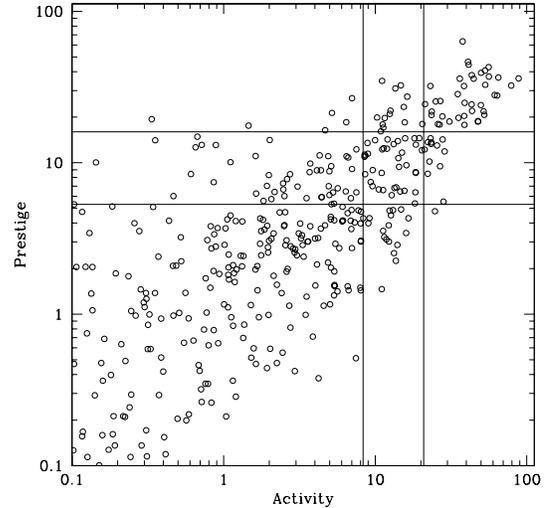

Fig. 19.— The Activity-Prestige diagram for all 928 individuals who received a PhD from a U.S. university in the five year period 1972–1976. The scaling of the axes and the horizontal and vertical lines are the same as in figure 18.

If we compare these two organizations directly against the ten faculty sample, as with the departments in table 4, SAO would have a score of 0.80 in SumProd (which would rank 7th) and 0.92 in Read10 (ranking 5th). Goddard would have scores of 0.27 in SumProd and 0.34 in Read10.

As federal civil service scientists can be at any level of seniority, and can be persons who would be viewed as technicians in academic departments, there is a bias in including these individuals in samples intended to measure the quality of publishing scientists. One way to compensate for this is to take a subset for the comparison. If we take the top half (probably too large a correction) we get SAO with a SumProd score of 1.44 (ranking 3rd) and a Read10 score of 1.69 (ranking 1st); Goddard would have 0.51 for SumProd (ranking 11th) and 0.73 for Read10 (ranking 8th). A more detailed analysis is beyond the scope of this paper.

With the productivity-percentile method of this section we have shown how to compare the average productivity of organizations independent of size, and we have indicated how to use these methods taking organization size and composition into account (by taking a fixed number of individuals or a subset of individuals for the representative sample).

### 5.2. Productivity–Productivity Diagrams

The Read-Cite diagrams of section 4 can obviously be used to compare groups of individuals simply by plotting the members of two organizations and comparing their positions in detail. To emphasize this point, we present another, similar diagram in figure 18.

Figure 18 shows citations vs Read10 for the tenured faculty sample, with the points crudely representing age, as noted in the caption. We have relabeled Read10 as Activity and (normalized) citations as Prestige to emphasize the expected use of this diagram. We have divided the tenured faculty sample in thirds, both by Read10 and citations, this is shown in the divisions in the diagram, persons with high activity and high prestige are in the top right of the diagram, for example.

Clearly the distributions of different groups in diagrams such as figure 18 can be analyzed and compared in a number of ways. As an example we show in figure 19 all 928 persons who received their PhDs from a U.S. university in the five year period 1972–1976, plotted on the same scale as in figure 18. As expected there are substantial differences between the age cohort and the faculties of the most prestigious organizations. The high activity–high prestige corner contains 26% of the faculty sample, but only 4% of the age cohort, and the low activity–low prestige corner contains 24% of the faculty sample but 80% of the age cohort.

### 5.3. Integrated Productivity–Productivity Diagrams

The methods in section 5.1 measure the average quality of organizations independent of size; but in frequent cases size matters. A good measure of the product of size and ability for an organization is the number of high ranking scientists in that organization. We choose to define high ranking as having a score which would be in the top third of the ten highest ranked faculties in the U.S., and we count these individuals using the activity and prestige measures Read10 and normalized citations.

Figure 20 shows the results for the eleven faculties ranked in table 4. The effect of size can be seen in the measures for U.C. Berkeley and U.C. Santa Cruz; they have exactly the same score (the symbols are offset on the plot for legibility) of 10 highly active and 9 highly prestigious people. As can be seen in table 4, the Berke-ley department is ranked about 50% above the Santa Cruz department by faculty quality, U.C. Santa Cruz is about 50% larger than U.C. Berkeley, thus in this figure they are equal. The small, high quality department at U. Minnesota has an (active, prestigious) score of (2,3) which would rank it below a number of larger departments.

TABLE 4. ASTRONOMY FACULTY RANK

| Institution | SumProd Rank | SumProd Score | Read10 Rank | Read10 Score | NRC Rank[a] |
|---|---|---|---|---|---|
| Princeton | 1 | 1.70 | 2 | 1.52 | 2 |
| CalTech | 2 | 1.46 | 4 | 1.28 | 1 |
| U.C. Berkeley | 3 | 1.32 | 3 | 1.44 | 3 |
| Harvard-Smithsonian | 4 | 1.21 | 1 | 1.60 | 4 |
| U.C. Santa Cruz | 5 | .83 | 7 | .82 | 6 |
| U. Hawaii | 6 | .82 | 5 | .87 | 11 |
| U. Arizona | 7 | .78 | 6 | .85 | 7 |
| U. Texas | 8 | .71 | 8 | .49 | 10 |
| U. Chicago | 9 | .57 | 11 | .39 | 5 |
| Cornell | 10 | .52 | 10 | .45 | 9 |
| M.I.T. | 11 | .46 | 9 | .47 | 8 |

[a]By perception of faculty quality (National Research Council 1995)

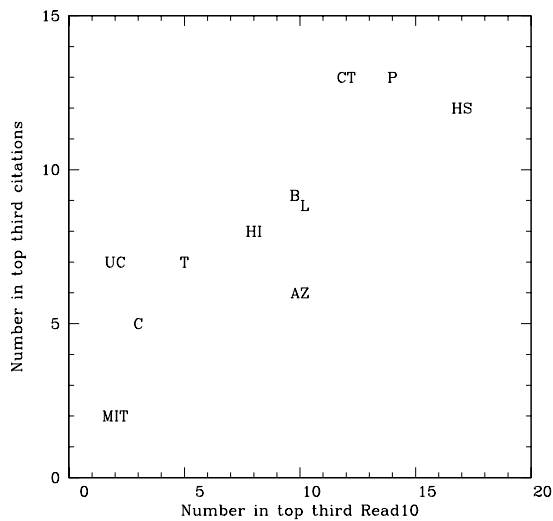

FIG. 20.— The number of faculty members from each organization who have Read10 scores in the top third of the ten faculty sample vs. the number whose citations rank in the top third. The meaning of the symbols is P: Princeton; CT: CalTech; HS: Harvard-Smithsonian; B: U.C. Berkeley; L: U.C. Santa Cruz (Lick observatory); HI: U. Hawaii; AZ: U. Arizona; T: U. Texas; UC: U. Chicago; C: Cornell; MIT: M.I.T.

It is also possible to measure the federal labs on the same scale as the universities using this method. The federal civil service scientists at the Smithsonian Astrophysical Observatory would have an (active, prestigious) score of (18,16), above any university in both measures, and the combined astrophysics labs at the Goddard Space Flight Center would have a score of (10,2).

## 6. CONCLUSIONS

Perhaps the most important new information to become available for bibliometric studies is the per article readership information. We now know how many times an article is read, where the reader is from, and "who" (as a unique cookie identifier, not as a name, which remains anonymous) the reader is. The existence of this information has great implications for the future of information retrieval and bibliometrics.

We expect the similarities and differences of reads and citations to become a central facet of bibliometric research. Whether emphasizing the similarities to achieve a statistically more stable result, as with our SumProd statistic, or emphasizing the differences to obtain a two dimensional view, as with the citations vs. Read10 diagram, the combination of the two measures of use substantially improves the capabilities of bibliometric measurement.

## 7. ACKNOWLEDGMENTS

It is a pleasure to thank I.I. Shapiro and J.P. Huchra for assistance in developing the tenured faculty sample; to thank R. Scheiber-Kurtz for extensive discussions, and for detailed reading and criticism of the text; to thank Edwin Henneken for reading the text; and to thank M.J. Geller for discussions, and for allowing the data in figure 19 to be shown in advance of publication.

MJK received the 2000 ISI/ASIST Citation Analysis Award for a preliminary version of section 4.

The ADS is supported by NASA under Grant NCC5-189.